\documentclass[aps,prb,superscriptaddress,twocolumn]{revtex4-1}

\usepackage{physics}
\usepackage{graphicx}
\usepackage{bm}
\usepackage{dsfont}
\usepackage{environ}

\begin{document}

\title{Monte Carlo study of the discontinuous quantum phase transition\\
in the transverse-field Ising model on the pyrochlore lattice}

\author{Patrick Emonts}
\affiliation{Institut   f{\"{u}}r  Theoretische Festk{\"{o}}rperphysik, JARA-FIT and JARA-HPC, RWTH Aachen University, 52056 Aachen, Germany}
\affiliation{Max-Planck-Institut f{\"{u}}r Quantenoptik, 85748 Garching, Germany}
\author{Stefan Wessel}
\affiliation{Institut   f{\"{u}}r  Theoretische Festk{\"{o}}rperphysik, JARA-FIT and JARA-HPC, RWTH Aachen University, 52056 Aachen, Germany}

\date{\today}

\begin{abstract}
The antiferromagnetic Ising model  on the pyrochlore lattice 
exhibits a quantum phase transition in an applied transverse field
from the low-field quantum spin-ice phase to the high-field polarized regime. Recent field-theoretical 
analysis and series expansion results indicate this to be a discontinuous, first-order transition. Here, we explore
this transition using quantum Monte Carlo simulations in order to assess this scenario and study
the thermodynamic properties in the vicinity of the quantum phase transition. For this purpose, we also
consider several variants of extended cluster-update schemes for the transverse-field Ising antiferromagnet on
frustrated lattices and compare their performance to the conventional bond-based algorithm 
for the transverse-field Ising model on the pyrochlore lattice.
\end{abstract}

\maketitle

\section{\label{introduction}Introduction}

\begin{figure}[t]
\centering
  \includegraphics[]{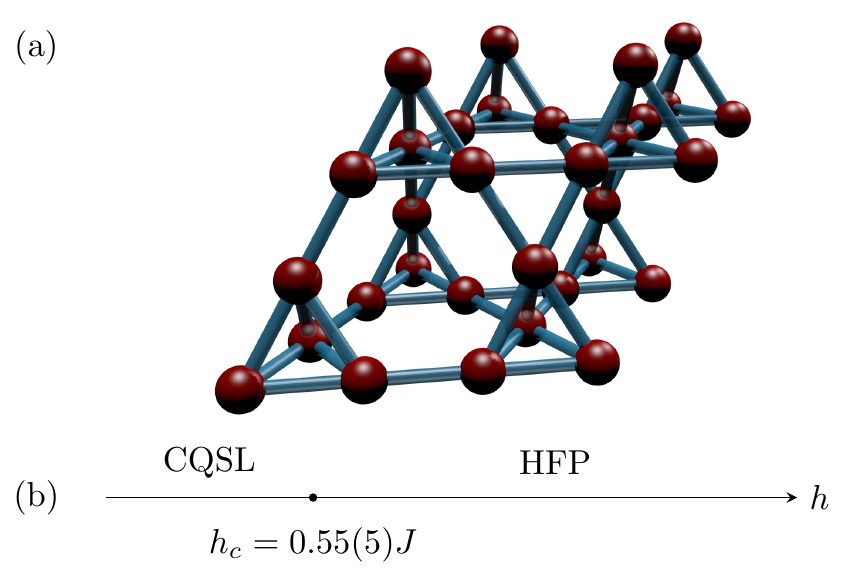} 
  \caption{(a) Illustration of a finite pyrochlore lattice 
  of a linear extent $L=2$ in each spatial direction, with $L^3=8$ unit cells and $N=4L^3=32$ lattice sites. 
  (b) Ground state phase diagram of the TFIM on the pyrochlore lattice with the discontinuous quantum phase transition at $h_c=0.55(5)J $ that separates the Coulomb quantum spin liquid (CQSL) phase from the high field polarized (HFP) regime.  The quoted value of $h_c$ is based on the QMC simulations reported in this work.}
\label{fig:model}
\end{figure}

Spin-ice materials such as $\text{Dy}_2\text{Ti}_2\text{O}_7$ or $\text{Ho}_2\text{Ti}_2\text{O}_7$
exhibit a thermodynamic  entropy at low temperatures that is comparable to the residual entropy of water ice~\cite{Harris97,Bramwell01,Ramirez99}.
The rare-earth magnetic ions  $\text{Dy}^{3+}$ and $\text{Ho}^{3+}$ in these materials are arranged
on a pyrochlore lattice, shown in Fig.~\ref{fig:model}, which consists of corner-sharing tetrahedra, in analogy to  the oxygen ions in water ice's wurtzite  structure. 
While the dominant interaction among the magnetic moments in spin-ice materials is dipolar~\cite{Siddharthan99,Ruff05,Melko04}, 
the characteristic residual entropy of spin ice is also displayed by the Ising model  with antiferromagnetic nearest-neighbor interactions on the pyrochlore lattice~\cite{Anderson56}.
 The dipolar character of the magnetic interaction in dipolar spin ice   eventually leads to the stabilization of a magnetically ordered low temperature state~\cite{Siddharthan99,Ruff05,Melko04}.  Instead, the highly frustrated classical Ising model  exhibits a highly degenerate ground space, formed by all spin configurations that satisfy the spin-ice rule; i.e., on each tetrahedron of the pyrochlore lattice, two Ising spins are present of either polarization~\cite{Anderson56}. The extensive degeneracy of these classical ground states leads to a finite residual entropy at zero temperature, $T=0$, much like the case of ideal water ice~\cite{Pauling35}. 

Quantum fluctuations and entanglement are introduced to such a classical Ising spin system upon adding transverse spin exchange interactions (leading to the XXZ model)~\cite{Hermele04}, or by applying a uniform magnetic field transverse to the quantization direction of the Ising spin exchange interactions~\cite{Lin12}.
It has been argued that, for non-Kramers degenerate rare-earth ions such as $\text{Dy}^{3+}$ and $\text{Ho}^{3+}$,
such effective transverse fields in the magnetic Hamiltonian derive from local electric field gradients~\cite{Savary17, Roechner17}.
While Ref.~\onlinecite{Savary17} considers disorder as a means of  inducing random transverse fields, the authors of Ref.~\onlinecite{Roechner17}
propose that strained non-Kramers spin-ice materials effect a uniform transverse field in the effective magnetic Hamiltonian. 
Under these conditions, the low-energy properties of the magnetic system may be described by the (uniform) transverse-field  Ising model (TFIM) on the pyrochlore lattice. 
This model is described by the Hamiltonian
\begin{equation}
  H=J\sum_{\langle i,j\rangle}\sigma_i^z\sigma_j^z-h\sum_i\sigma_i^x,
  \label{eq:model}
\end{equation}
where the Pauli matrices $\sigma^\alpha_i$ act on  the spin-1/2 moment on lattice site $i$ of the pyrochlore lattice, $J>0$ denotes the nearest-neighbor longitudinal antiferromagnetic Ising coupling strength and $h$ the strength of the  transverse field. 
Within the proposal of Ref.~\onlinecite{Roechner17}, the effective field strength $h$ is controlled by the  magnitude of the strain applied to the  spin-ice material.
Depending on the ratio $h/J$, the ground state of the Hamiltonian $H$ furthermore exhibits two distinct phases~\cite{Savary17, Roechner17}; cf.  the phase diagram in Fig.~\ref{fig:model}.
In the low-field regime ($h\ll J$), quantum fluctuations were argued in  Ref.~\onlinecite{Savary17}  to stabilize a quantum spin-ice phase, which is characterized by a Coulomb quantum spin liquid (CQSL)  with an emergent phonon soft mode and gapped electric and magnetic excitations~\cite{Hermele04, Shannon12}.
It has furthermore been concluded, based on both field-theoretical analysis and series expansion calculations, that the transition out of this CQSL phase into the high-field polarized (HFP) region ($h\gg J$), i.e., without symmetry breaking on either side of the transition, proceeds via a discontinuous first-order quantum phase transition~\cite{Savary17, Roechner17}. 
Within the field-theory formulation, this transition is characterized as a confinement transition, induced by the condensation of bosonic monopoles, which can be described by a complex scalar field that is coupled to a U(1) gauge field.  
The mathematical structure of this quantum field theory corresponds  to that of the Ginzburg-Landau theory of a superconductor in four dimensions. 
From the renormalization group analysis of this theory, it is known that in such a system the confinement transitions is in fact first order~\cite{Halperin74}. 
Within this scenario, the series expansion calculations in Ref.~\onlinecite{Roechner17}  provide an estimate for the field strength of $h_c\approx 0.6 J$ at the corresponding  quantum phase  transition in the TFIM on the pyrochlore lattice. 
This value is close to the estimate $h_c=0.55(5)J $ that we obtain from our  analysis in Sec.~\ref{sec:results}; cf.  also the phase diagram in Fig.~\ref{fig:model}. 

While the series expansion calculations were performed in the thermodynamic limit, other computational approaches to examine this transition, such as quantum Monte Carlo (QMC) simulations, which we consider here, instead require the analysis of finite-size systems. 
Within such finite-size approaches, the identification of a first-order quantum phase transition in the TFIM is peculiar, since this transition does not persist as a level-crossing transition in finite systems; this is a consequence of the noncrossing theorem, i.e., no level crossing can occur in the ground state of the TFIM as only a single parameter of the Hamiltonian (here, the transverse field strength $h$), is varied~\cite{Ruskai02}. 
Instead, the nondegenerate ground state of the quantum model in a finite system evolves continuously with $h$, such that discontinuous behavior may emerge only in the thermodynamic limit. 
The nonzero excitation gap of a finite-size system, closing at the quantum phase transition point in the thermodynamic limit, sets a corresponding low-temperature scale, above which the emerging discontinuous behavior
can effectively be probed.
As we discuss below, we obtain indications for such an emerging discontinuous behavior within an appropriate low-temperature regime of the TFIM on finite pyrochlore lattices that are accessible to QMC simulations.

One of the characteristics of the Hamiltonian in Eq.~(\ref{eq:model}) is the high geometric frustration from  the antiferromagnetic Ising interactions among the spins on the pyrochlore lattice due to the corner sharing tetrahedrons. In contrast to many other frustrated quantum spin systems, for which QMC simulations suffer from a severe sign problem, one can, however, still simulate the Hamiltonian in Eq.~(\ref{eq:model}) using sign-problem-free QMC algorithms.
Namely, 
in order to perform the QMC simulations for this highly frustrated quantum spin system, we used the stochastic series expansion (SSE) formulation~\cite{Sandvik91,Sandvik99,Sandvik11}, employing different previously formulated TFIM Hamiltonian decompositions and cluster-update schemes for this purpose~\cite{Sandvik03,Biswas16}. 
Furthermore, we  devised two additional cluster-update schemes that are adapted to the pyrochlore lattice. 
We compare in particular the efficiency of these various simulation schemes: Namely, due to the  
highly frustrated interactions, the QMC simulations remain challenging in the low-temperature regime, as detailed further below. 
Based on these algorithms, we  then  characterize the emerging low-temperature first-order phase transition of the TFIM model on the pyrochlore lattice. 

The further layout of this paper is as follows: 
In Sec.~\ref{sec:methods}, we outline the SSE QMC algorithm for the TFIM and discuss previously proposed as well as further pyrochlore-lattice-adapted decomposition and cluster-update schemes.  
We then present in Sec.~\ref{sec:results} our  results for the  low-temperature properties of the Hamiltonian in Eq.~(\ref{eq:model}) within  the field-induced transition region.
Finally, we draw conclusions in Sec.~\ref{sec:conclusions}.

\section{\label{sec:methods}QMC Methods}
The QMC methods that we used in our calculations are based on the SSE formulation~\cite{Sandvik91,Sandvik11} of the quantum partition function 
\begin{align}
  Z&=\mathrm{Tr} \:e^{-\beta H}=\sum^{\infty}_{n=0} \frac{\beta^n}{n!} \sum_{\alpha} \bra{\alpha}(-H)^n\ket{\alpha}
  \label{eq:mc_partition_taylor}
\end{align}
of the system at an inverse temperature $\beta=1/T$ (we set $k_B=1$). 
Here, $\ket{\alpha}$ denotes the computational basis that is used to express the trace in $Z$. 
In the following, we will always use the  basis of the eigenstates of the Ising coupling term, i.e., each basis state $\ket{\alpha}=\ket{\sigma_1^z,...,\sigma_N^z}$ is specified by the eigenvalue $\pm 1$ of $\sigma^z_i$ for each spin $i=1,...,N$, where $N$ is the total number of spins on the finite pyrochlore lattice. 
In order to formulate a numerical simulation scheme, one proceeds by decomposing the Hamiltonian $H$ into a sum of diagonal and off-diagonal terms, with each such term containing operators acting on only a small cluster of sites from the full lattice. 
For a Hamiltonian with nearest-neighbor interactions, the smallest feasible clusters consist of local (onsite) and  nearest-neighbor (two-site) bond Hamiltonians on the considered lattice. 
A QMC algorithm based on such a bond decomposition for generic TFIM Hamiltonians with  two-site exchange interactions of arbitrary sign and extent on any lattice was formulated by Sandvik in Ref.~\onlinecite{Sandvik03}. 
The same paper  also presents a cluster-update scheme for the sampling of the corresponding QMC configuration space. This algorithms is denoted as B-SSE in this paper. 
We found that the B-SSE algorithm performs efficiently  also  for the TFIM on the pyrochlore lattice in the vicinity of the quantum phase transition.  
We thus consider this algorithm in the following and  provide a short review in  Sec.~\ref{sec:bond}. 

Alternatively to the B-SSE algorithm, Biaswas, Rakala and Damle proposed a variant SSE simulation scheme for the TFIM that is based on a decomposition of the Hamiltonian into local as well as triangular (three-site) plaquettes~\cite{Biswas16}. 
In the following, we denote this algorithm as P-SSE. 
While in its original formulation the P-SSE algorithm was examined and applied to the TFIM on the triangular lattice, the authors of Ref.~\onlinecite{Biswas16} suggest that a triangular decomposition scheme can also be employed for other lattice structures that allow for a decomposition into a set of triangular plaquettes. 
We thus examine the application of the P-SSE algorithm of Ref.~\onlinecite{Biswas16} to the TFIM on the pyrochlore lattice in Sec.~\ref{sec:plaquette}. 
Based on our analysis, we find that the P-SSE algorithm from Ref.~\onlinecite{Biswas16} is not reliable on the pyrochlore lattice. 
We thus devised a refined plaquette-based algorithm for the pyrochlore lattice, which uses  a grouped plaquette cluster-update scheme, and which we denote as G-SSE in the following. 
This algorithm exhibits superior performance as compared to the P-SSE method, and  we thus present it in Sec.~\ref{sec:grouped}. 
Finally, given that the pyrochlore lattice can naturally be decomposed into tetrahedra (four-site clusters), we also devised a tetrahedron-based decomposition scheme and an QMC update method of the TFIM on the pyrochlore lattice based on this tetrahedron decomposition. 
This algorithm will be denoted as T-SSE and will be presented in Sec.~\ref{sec:tetra}.

Within all the QMC algorithms that we employed for our investigation, we use  update schemes that allow for  independent flips of all clusters. These are constructed during each QMC update step, like in  the Swendsen-Wang~\cite{Swendsen87} cluster-update scheme for the classical Ising model.  We note that alternative simulation schemes for the TFIM on the pyrochlore lattice may also be considered. In particular, one can devise single-cluster construction schemes that resemble the classical Wolff algorithm~\cite{Wolff89} to control the cluster growth. 
An implementation of such a single-cluster approach within the bond-decomposition scheme  
was, however, not found to  lead to  better performance in Ref.~\onlinecite{Humeniuk18}. Furthermore, one may also apply a standard Trotter-Susuki decomposition~\cite{Suzuki12,Lin12} scheme to the quantum partition function of the TFIM and obtain a path-integral formulation for this model on the pyrochlore lattice. For an efficient sampling, one may extend the membrane-algorithm proposed by Henry and Roscilde~\cite{Henry14} for the TFIM on the checkerboard lattice to the case of the pyrochlore lattice, combined with an extension of the classical cluster updates described in  Refs.~\onlinecite{Kandel92, Barkema98, Newman99} to the pyrochlore lattice~\cite{Melko04}.  We did not follow these alternative directions further, but  they may be useful for exploring the  low-field regime of the TFIM on the pyrochlore lattice, in order to confirm the predictions from the CQSL scenario, similarly to the recent study of the XXZ model on the pyrochlore lattice~\cite{Huang18}. 
Here, however, we concentrate on the field-induced quantum phase transition at $h_c/J\approx 0.55(5)$, using the SSE algorithms presented in the remainder of this section. 

\subsection{\label{sec:bond}Bond decomposition scheme (B-SSE)}
The B-SSE algorithm for  the TFIM Hamiltonian was introduced in Ref.~\onlinecite{Sandvik03} and allows for the simulation of TFIMs with arbitrary interactions within the SSE approach (cf. also Ref.~\onlinecite{Biswas16} for a detailed description of this approach). 
For the TFIM in Eq.~\eqref{eq:model}, we introduce three kinds of partial Hamiltonian terms, 
\begin{align}
  H_{i,0}&=h(\sigma_i^++\sigma_i^-), \label{eq:mc_decomp_flip}\\
  H_{i,1}&=h\mathds{1}_i, \label{eq:mc_decomp_transverse}\\
  H_{b,2}&= J-J\sigma_{i(b)}^z\sigma_{j(b)}^z, \label{eq:mc_decomp_ising}
\end{align}
where $\sigma_i^\pm=\frac{1}{2}(\sigma_i^x\pm i \sigma_i^y)$.
Here, $i=1,...,N$ extends over the lattice sites, and 
 $b=1,...,N_b=3N$ over the  nearest-neighbor bonds on the pyrochlore lattice, where bond $b$ connects the two sites $i(b)$ and $j(b)$. 
 For a finite pyrochlore lattice of linear extent $L$ in terms of unit cells, the number of lattice sites $N=4L^3$, cf. Fig.~\ref{fig:model}, and we employ periodic boundary conditions in all three spatial directions (the number of unit cells is equal to $L^3$). 
 Finally, $\mathds{1}_i$ in Eq.~(\ref{eq:mc_decomp_transverse}) denotes the identity operator on the $i$th spin.  
Since there is only a single Hamiltonian term for a given  bond $b$, the second, numerical index of 2 on $H_{b,2}$ is actually redundant, but is kept here for convenience when formulating the SSE representation  of the partition function in the following. 
Upon summation,
\begin{align}
  H=-\sum_i(H_{i,0}+H_{i,1})-\sum_b H_b+C,
  \label{eq:mc_summing_convention}
\end{align}
the original Hamiltonian $H$ is reproduced  up to a constant, with $C=N(h+3J)$ for the B-SSE algorithm. 
The operators $H_{i,0}$ are the off-diagonal spin-flip operators that couple the spin at site $i$ to the transverse field. 
The other operators are diagonal in the Ising basis of the $\ket{\alpha}$ states.  For the construction of the cluster-update scheme, it is crucial to note that the nonzero matrix elements of both local site Hamiltonians $H_{i,0}$ and  $H_{i,1}$ are equal to $h$. Similarly, the matrix elements of the 
 bond Hamiltonians $H_b$ vanish if they are acting on a state with parallel spins on sites $i(b)$ and $j(b)$; for antiparallel spins, the matrix elements equal $2J$. 
Inserting the above decomposition of $H$ in terms of site and bond operators into the high-temperature expansion of $Z$, one obtains an expression for $Z$ as a sum over operator sequences:
\begin{align}
  Z=\sum_\alpha \sum_{n=0}^\infty \sum_{S_n} \frac{\beta^n}{n!}\langle \alpha | \prod_{q=1}^n H_{l(q),k(q)} | \alpha \rangle.
  \label{eq:Z_fixedlength}
\end{align}
In this expression, each operator sequence $S_n=([l(1),k(1)],...,[l(n),k(n)])$ is specified by a string of index pairs that specify the above partial Hamiltonian operator terms in terms of its locations on the lattice [i.e., $l(q)$ can be a site or a bond index] and the kind  of operator [i.e., $k(q)$ can take on values $0$ and $1$ for a site operator, and equals $2$ for a bond operator]. 
For a given finite system size and inverse temperature $\beta$, the expansion order $n$ can be constrained  to a sufficiently large value
$n<\Lambda\sim N\beta  $, without introducing a systematic error in the QMC sampling~\cite{Sandvik91, Sandvik03}. The appropriate value of $\Lambda$ is determined within the thermalization phase of the QMC simulation~\cite{Sandvik91, Sandvik03}.
In order to formulate an QMC sampling scheme of the  truncated partition function, one introduces one more operator, $H_{0,0}=\mathds{1}$, i.e., the identity. This extension allows one to express the truncated partition function as a sum over fixed-length operator sequences $S_\Lambda=([l(1),k(1)],...,[l(\Lambda),k(\Lambda)])$ that can include also identity $H_{0,0}$ operators:
\begin{align}
  Z=\sum_\alpha \sum_{S_\Lambda} \beta^{n(S_\Lambda)}\frac{[\Lambda-n(S_\Lambda)]!}{\Lambda!}\langle \alpha | \prod_{q=1}^\Lambda H_{l(q),k(q)} | \alpha \rangle. 
  \label{eq:Z_fixedlength_id}
\end{align}
Here, the effective expansion order $n(S_\Lambda)$ denotes the number of non-$H_{0,0}$ operators contained in the operator sequence $S_\Lambda$.
The factor $[\Lambda-n(S_\Lambda)]!/\Lambda!$ results from accounting for the number of possibilities of inserting the identity operators at arbitrary positions of the original operator strings. After the initialization of the QMC configuration by, e.g., a randomly chosen basis state $\ket{\alpha}$ and an  operator string $S_\Lambda$ that consists of $H_{0,0}$ operators only [i.e., $n(S_\Lambda)=0$], 
the QMC sampling of $Z$ is  performed by the iteration of the following two consecutive update steps: the (i) diagonal update, during which the effective  expansion order $n(S_\Lambda)$  varies, while the base state $\ket{\alpha}$ remains fixed, and (ii) a cluster-update step,  at a fixed value of $n(S_\Lambda)$, that can effect a new base state, and by which diagonal and off-diagonal operators may be interchanged. 

The diagonal update proceeds by traversing the operator string from $q=0$ to $q=\Lambda$. If the $q$th operator is an identity $H_{0,0}$ operator, a Monte Carlo update of the operators sequence is proposed that replaces $H_{0,0}$ by a randomly chosen diagonal Hamiltonian term. 
This may  either be a local operator of type $H_{i,0}$, or an Ising operator $H_{b,2}$ on a bond of the pyrochlore lattice. 
In reverse, if  the $q$th operator is a diagonal Hamiltonian, we propose to replace this operator by the identity operator $H_{0,0}$. 
The corresponding update probabilities depend explicitly on  the matrix elements of the involved operators, and can be  obtained from Eq.~\ref{eq:Z_fixedlength_id} using the detailed balance condition. 
We refer to Refs.~\onlinecite{Sandvik03, Biswas16} for further details. 

\begin{figure}[t]
  \includegraphics[]{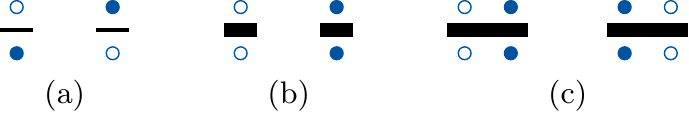} 
  \caption{
    Illustration of the three vertices used in the B-SSE algorithm: (a) spin-flip vertex [Eq.~\eqref{eq:mc_decomp_flip}], (b) transverse-field (constant) vertex [Eq.~\eqref{eq:mc_decomp_transverse}], and (c)~Ising  vertex [Eq.~\eqref{eq:mc_decomp_ising}]. The vertices in (a) and (b) act on a single spin.  The Ising vertex (c) acts on two spins (two small dots are attached). Open~(filled) dots symbolize an Ising spin pointing down (up).}
  \label{fig:mc_vertices}
\end{figure}

We next outline the B-SSE cluster-update scheme from Ref.~\onlinecite{Sandvik03}. 
For this purpose, it is convenient to represent a given QMC configuration in terms of vertices. 
For the site terms $H_{i,0}$ and $H_{i,1}$, these vertices consist of one incoming state (forming a leg to this vertex ) and one outgoing state (leg), as depicted in Fig.~\ref{fig:mc_vertices}. 
Correspondingly, the bond Hamiltonians $H_b$ are represented by vertices with two incoming  legs and two outgoing legs, which are also shown in Fig.~\ref{fig:mc_vertices}. 
The figure only shows those local configurations, for which the corresponding matrix elements are finite. 
These vertex states are thus the allowed QMC vertex configurations, i.e., these are contained in QMC configurations that give finite contributions to the partition function. 
Each valid QMC configuration, given in terms of a basis state $\ket{\alpha}$ and an operator sequence $S_{\Lambda}$, can now be represented by  a doubly linked list of vertices (the identity operators $H_{0,0}$ are ignored for this purpose)~\cite{Sandvik99,Sandvik03,Sandvik11}. 
Namely, the incoming and outgoing legs of the vertices are connected according to the order of the operators in the operator sequence and the action of each such operator on the corresponding spins. 

The cluster-update scheme now proceeds by constructing clusters of vertex legs in such a way that each leg of each vertex is eventually contained within a unique cluster. By the form of the Hamilton decomposition, one is then ensured that the QMC weight  does not change if one flips the spin states on all the legs that are connected within a given cluster. Thus, after the construction of all the clusters, we propose for each such cluster to flip the spins on all the legs that belong to this cluster with a probability of $1/2$. This corresponds to the Swendsen-Wang cluster-update scheme for the classical Ising model~\cite{Swendsen87}. 
Such a cluster-update scheme effects global updates of the QMC configuration, changing  both the base state as well as the content of  the operator sequence, since after a cluster flip the configuration of the spins on the legs of a two-leg vertex  corresponds to an  operator of a different  kind than before. 

\begin{figure}[t]
  \centering
  \includegraphics[]{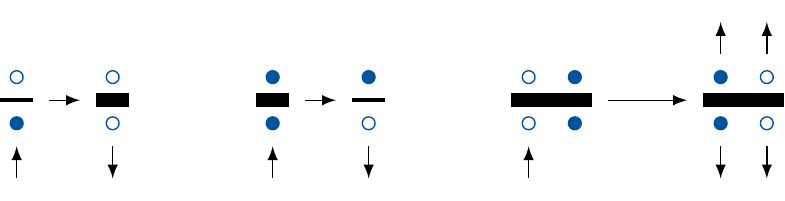}
  \caption{Illustration of the cluster building rules for the B-SSE algorithm.
    The cluster building rules are shown by the marked entrance and exit legs of the vertices.
    The change of the local spin configuration upon an accepted cluster flip is also shown. 
  }
  \label{fig:bondupdate}
\end{figure}

In detail, the clusters are constructed as follows: 
We consider each leg of each vertex consecutively. 
If this leg has not been already added  to a cluster, we start to construct a new cluster by adding this leg onto a stack. 
As long as this stack is not empty (in particular after just adding the starting leg), we pull the top leg from the stack, add it to the cluster, and then 
attempt to grow the cluster using the following rules: 
(1) If the current leg belongs to a two-leg vertex, we follow the link from this leg in the doubly linked list of connected vertices. 
If the thereby reached leg does not yet belong to any cluster and is not already on the stack, we add it to the stack. 
(2) If on the other hand, the current leg belongs to a four-leg vertex, we also add the other three legs of this vertex onto the stack. 
We then follow the doubly linked list from each of these four vertex legs, and also add those reached legs onto the stack that have not been already added to any cluster or the stack. 
These local rules are illustrated in Fig.~\ref{fig:bondupdate} for the three different kinds of vertices in a specific spin configuration.
In this figure, an arrow  pointing into a vertex indicates the incoming leg on that vertex,  which was pulled off the stack most recently, while the outpointing arrows indicate the exit legs on that vertex, which are followed in the doubly linked list for the  consideration of further legs according to the above rules. 

After having followed rule (1) or (2), we pull off the next leg from the stack. 
This process continues until the stack is eventually empty again. 
We have thereby constructed a new cluster. 
As long as there still remain legs that have not been added to any cluster so far, we continue to construct another cluster. 
Eventually, all spins will belong to one of the constructed clusters, and we can then propose the Swendsen-Wang cluster flip for each cluster as already mentioned. In practice, the decision whether a newly constructed cluster will be flipped can be made already at the beginning of the new cluster construction, and thus the leg updates can be performed on the fly during the cluster construction. 

\begin{figure}[t]
\centering
\includegraphics[]{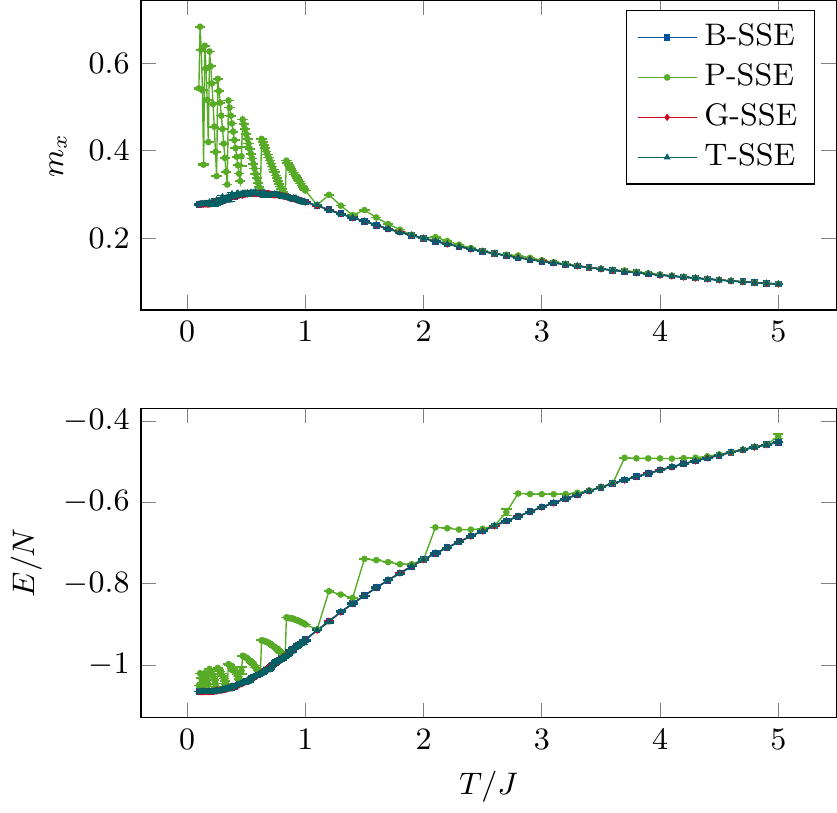}
\caption{Temperature dependence of the transverse magnetization $m_x$ (top panel) and the energy $E$ (bottom panel), of the TFIM on the pyrochlore lattice at $h=0.5J$, as obtained for a $L=4$ pyrochlore lattice using the various considered QMC algorithms.
In all cases, we performed $10^6$  thermalization sweeps followed by another $10^6$ measurement steps.}
\label{fig:comp_temp_scan_4x4x4_h_0_5}
\end{figure}

As an example  of the application of the B-SSE algorithm to the TFIM on the pyrochlore lattice, we show in Fig.~\ref{fig:comp_temp_scan_4x4x4_h_0_5} the temperature dependence of both the energy  $E=\langle H \rangle$ and the transverse magnetization 
\begin{equation}
m_x=\frac{1}{N}  \left\langle \sum_{i=1}^N \sigma_i^x\right\rangle
\end{equation}
at a transverse-field strength of $h=0.5J$, close to the anticipated quantum phase transition, for a lattice with $L=4$, containing $N=4L^3=256$ lattice sites. 
Within the SSE approach, these quantities are readily obtained as  
$E=-\langle n \rangle_\mathrm{MC}/\beta+C$ and $m_x=\langle n_0 \rangle_\mathrm{MC}/(N\beta h)$, respectively, where $\langle n \rangle_\mathrm{MC}$ denotes the Monte Carlo expectation value of the expansion order, and $\langle n_0 \rangle_\mathrm{MC}$ the Monte Carlo expectation value of the number $n_0$ of transverse field $H_{i,0}$ operators $(i=1,...,N)$ contained in the sampled operator sequences~\cite{Sandvik91}.
We find  that within this parameter regime of the transverse-field strength, we can probe the thermodynamic properties down to temperatures of about $T\approx 0.01 J$. 
In the same figure, we show in addition to the results obtained from the B-SSE algorithm also the results that we obtained using several other decompositions and corresponding cluster-update schemes, which we will discuss next. 

\subsection{\label{sec:plaquette} Plaquette decomposition scheme (P-SSE)}
The B-SSE algorithm outlined in the previous section is based on a decomposition of the Hamiltonian into single-site and bond terms, i.e., the basic building blocks of the TFIM Hamiltonian, and it can thus be directly applied  to the pyrochlore lattice case. 
However, the performance of the B-SSE algorithm in sampling within the low-temperature regime is
impaired by the large degeneracy 
that characterizes the classical antiferromagnetic Ising model on the pyrochlore lattice due to the geometric frustration on this lattice, in particular in the low-field regime, where the frustrating Ising exchange prevails.
One attempt to alleviate the sampling for such highly frustrated spin models is based on contriving other decomposition schemes of the TFIM Hamiltonian. The authors of Ref.~\onlinecite{Biswas16} present an alternative algorithm for lattices composed  of triangular units, which they suggest can be applied also to the pyrochlore lattice. This P-SSE algorithm is based upon a decomposition of the Hamiltonian in terms of triangular plaquettes.
The  motivation behind such a decomposition stems from the observation that the bond-based algorithm fails to sample from  the low-energy  states of the minimally frustrated Ising model configurations on the triangular lattice, i.e., states with one frustrated bond per triangle~\cite{Biswas16}. By instead decomposing the Hamiltonian in terms of  triangular plaquettes, which  contain this local frustration explicitly, one may alleviate this difficulty and thus ensure a more effective sampling. Based on this insight, one thus decomposes the Ising interaction of the Hamiltonian into the elementary triangular plaquettes (each formed by three neighboring spins that are connected into a triangular cluster) of the considered lattice. The new Ising operators $H_{p,2}$, which replace the bond Hamiltonians $H_{b,2}$ from the bond-decomposition scheme, are defined as
\begin{equation}
H_{p,2}= \frac{3}{2}J - \frac{J}{2} \left(\sigma^z_{i(p)} \sigma^z_{j(p)} +  \sigma^z_{j(p)} \sigma^z_{k(p)} + \sigma^z_{k(p)} \sigma^z_{i(p)}\right),
\end{equation}
where $i(p)$, $j(p)$, and $k(p)$ denote the three sites that form the considered triangular plaquette $p$ of the pyrochlore lattice.
The prefactor $1/2$ of the Ising coupling terms reflects the fact that each bond on the pyrochlore lattice 
is shared by two elementary triangular plaquettes (cf. Fig.~\ref{fig:model}). The total number of elementary triangular plaquettes on a system of  linear extent $L$ with  $N=4L^3$ sites is equal to $N_p=8L^3$.
Being diagonal operators in the computational basis, these Ising operators can be inserted into, as well as removed from, the operator string within the diagonal update, which thus proceeds similarly as for the bond-decomposition scheme. 
The insertion and deletion probabilities can be found in Ref.~\onlinecite{Biswas16}.  

Before we describe the cluster-update scheme from Ref.~\onlinecite{Biswas16}, which is based on the plaquette decomposition, a few further definitions need to be introduced. 
Prior to performing a cluster-update step, one singles out one of the three sites $i(p)$, $j(p)$ or $k(p)$ as a {\it privileged} site
for each of the triangular plaquettes  that correspond to  the operators $H_{p,2}$. 
The triangular lattice consists  of three sublattices A, B, C, so that a ``natural'' assignment of the privileged sites can be performed that is globally consistent along all triangular plaquettes of the lattice (namely, consider all the sites from sublattice A as privileged, etc.)~\cite{Biswas16}. For the pyrochlore lattice, such a globally consistent assignment of the privileged site to each triangular plaquette is, however, not possible. This can be seen already when considering the four triangles on a single tetrahedron.  Hence, in order to apply the cluster-update scheme from Ref.~\onlinecite{Biswas16} to the pyrochlore lattice, we instead independently picked out one of the three sites $i(p)$, $j(p)$, and $k(p)$ as the privileged site for each triangular plaquette $p$. This procedure of locally assigning the privileged spin can still be performed in various ways (e.g., at random or according to a given order).  However, we did not observe significant differences in the obtained QMC data from  using  different such assignment schemes. 

Furthermore, for a given state of the spins at the legs of an $H_{p,2}$ operator vertex, we follow Ref.~\onlinecite{Biswas16} to 
denote as the {\it minority} spin the spin (out of the three spins that form the triangular plaquette of this operator) that is oriented oppositely to the other two spins. Note, that maximally frustrated configurations,  in which all three spins are parallel, do not appear as they have zero weight. The other two spins are called the {\it majority} spins of this vertex. 
The terms privileged site and majority or minority spin are entirely independent concepts: 
the privileged site is defined without reference to  the actual spin configuration of a given vertex.

\begin{figure}[t]
  \centering
  \includegraphics[]{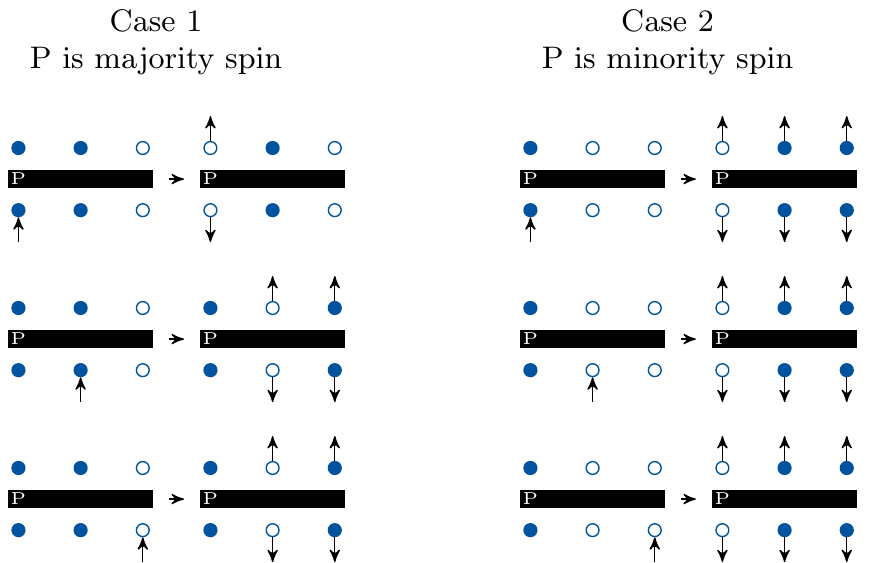}
  \caption{Illustration of the cluster building rules for the P-SSE  algorithm.
    The cluster building rules are shown by the marked entrance and exit legs of the vertices.
    The change of the local spin configuration upon a  cluster flip are also shown. 
    The privileged site is marked by P.
  }
  \label{fig:plaquetteupdate}
\end{figure}

Before starting the actual cluster construction, one finally identifies, for each $H_{p,2}$ operator vertex in the operator sequence, whether the spin at the assigned privileged site on that triangle is a majority or a minority spin for the local spin configuration of the legs of that vertex. 
The cluster construction rules (see also Fig.~\ref{fig:plaquetteupdate}) upon entering a $H_{p,2}$ operator then depend on whether the spin at the privileged site is a majority or a minority spin. 
This distinction ensures that no maximally frustrated triangular  plaquettes (with zero weight) are generated. 
If the privileged site is a majority spin and the entrance leg corresponds to the privileged site, then the other leg that corresponds to the privileged site is added to the stack and taken as an exit leg along with the entrance leg.  
If the privileged site is a majority spin and the entrance leg corresponds to one of the two nonprivileged sites, all legs corresponding to the nonprivileged sites are added to the stack and taken as exit legs. 
If, on the other hand, the privileged site is a minority spin, all six legs are added to the stack and taken as exit legs. The cluster building rules for the two-leg vertices, i.e., single-site vertices, stay the same as described in the B-SSE cluster-update scheme. 
The rules for the new quantum cluster-update scheme are summarized exemplarily for the first site as privileged site in Fig~\ref{fig:plaquetteupdate}.  

To test the applicability of the P-SSE algorithm to the TFIM on the pyrochlore lattice, we first compare  results obtained using this algorithm to those obtained from the B-SSE algorithm for the $L=4$ system at $h=0.5J$; cf. Fig.~\ref{fig:comp_temp_scan_4x4x4_h_0_5}. 
While at large temperatures $T \gtrsim 2 J$, the results for $m_x$ from the P-SSE algorithm appear consistent with those from the B-SSE algorithm, we observe convergence problems at lower temperatures, both in the transverse magnetization $m_x$ and the energy.
Since the ergodicity of the QMC algorithm is  related to the design of the cluster-update procedure, we modified the cluster-update scheme of the P-SSE algorithm in order to restore its convergence also on the pyrochlore lattice.
This new algorithm is motivated by minimizing the number of updates that flip the whole operator at once, i.e., exiting on all legs.
The off-diagonal update steps is based on a grouping procedure (hence the name G-SSE).

\subsection{\label{sec:grouped} Grouped plaquette update scheme (G-SSE)}
\begin{figure}[t]
 \centering
 \includegraphics{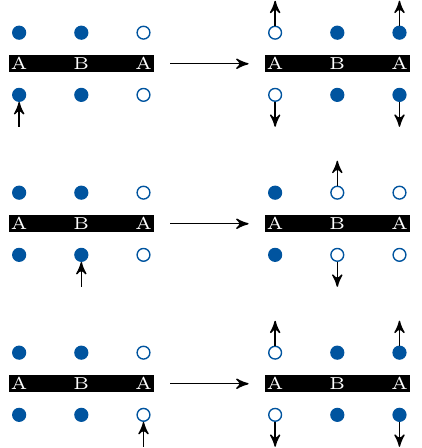}
 \caption{Illustration of the cluster building rules for the G-SSE algorithm.
    The cluster building rules are shown by the marked entrance and exit legs of the vertices.
    The changes of the local spin configuration upon an accepted cluster flip are also shown. 
  }
  \label{fig:gplaquette_offdiagonal_update}
\end{figure}

The grouped plaquette update scheme of the G-SSE algorithm is a variant of the plaquette-based algorithm of Ref.~\onlinecite{Biswas16}, and  converges to the actual thermodynamic properties on the pyrochlore lattice.
This results from an adaptation of the cluster-construction rules to the geometry of pyrochlore lattice while keeping a decomposition in terms of triangles. 
The partitioning of the Hamiltonian and the diagonal update are thus the same as in the original  P-SSE algorithm of Ref.~\onlinecite{Biswas16}.
The modified cluster-update avoids exiting on all six legs of an $H_{p,2}$ vertex, and reduces the number of attempts of changing the majority spin as much as possible. Therefore, it tends to conserve the spin-ice configurations of the pyrochlore lattice.
Instead of using minority and majority updates based on the concept of  privileged sites, we now separate the sites within a given triangular plaquette for an operator $H_{p,2}$  into  two groups (denoted A and B). In particular, we assign the minority spin to group A. Furthermore, one of the two majority spins is also assigned to group A, while the second majority spin is assigned to group B. This assignment is performed randomly for each $H_{p,2}$ operator in the operator string, prior to performing the cluster construction. 
Thus, for each operator $H_{p,2}$, there are always two sites in group A and one site in group B.
The original cluster construction at the $H_{p,2}$ vertices is now replaced by the following simple rule:
if we enter a $H_{p,2}$ vertex on a leg that  corresponds to a site from group A (B), 
we also exit on all legs that correspond to the sites of the same group A (B).
This  cluster building rule is illustrated for an exemplary  spin configuration in Fig.~\ref{fig:gplaquette_offdiagonal_update}.
We finally note that one Monte Carlo step of the G-SSE algorithm consists of one diagonal and one cluster-update step. This fact makes the  B-SSE algorithm and the G-SSE plaquette algorithm directly comparable in terms of update steps.

\begin{figure}[t]
 \centering
 \includegraphics[]{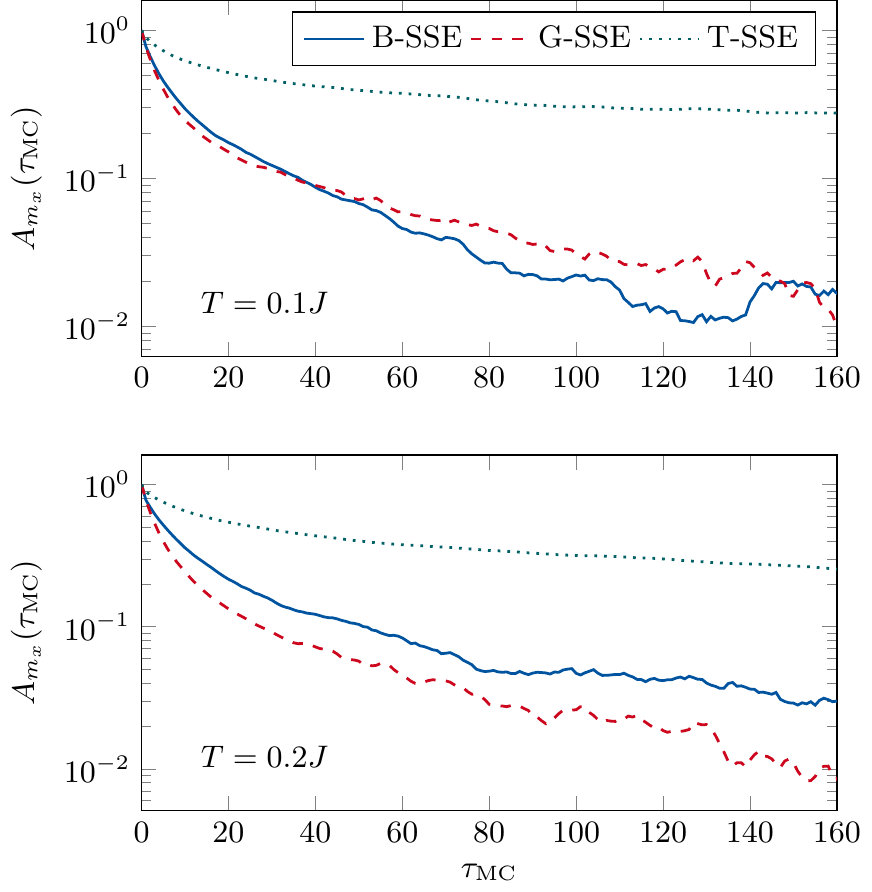}
 \caption{Comparison of the autocorrelation function $A_{m_x}(\tau_\mathrm{MC})$ of the transverse magnetization $m_x$ using the
 B-SSE, the G-SSE, as well as the T-SSE  algorithm for $T=0.1J$ (top panel) and $T=0.2J$ (bottom panel), as measured from QMC simulations of the TFIM on the pyrochlore lattice for  $L=6$ and $h=0.3J$. Here, $\tau_\mathrm{MC}$ denotes the number of Monte Carlo steps.  }
  \label{fig:comp_auto}
\end{figure}

In Fig.~\ref{fig:comp_temp_scan_4x4x4_h_0_5}, we compare the results for the energy $E$ and the transverse magnetization $m_x$ as  obtained from the G-SSE algorithm to the results from the previous two approaches.
We find that in contrast to the P-SSE algorithm, the G-SSE algorithm performs well down to low temperatures and also returns results that are consistent with the B-SSE method.
Upon examining the performance of the B-SSE vs~the G-SSE scheme, we observed that both lead to  overall  similar autocorrelations times. 
To quantify this more specifically, we show a comparison of the autocorrelation function $A_{m_x}(\tau_\mathrm{MC})=\langle m_x(\tau_\mathrm{MC}) m_x(0)\rangle_\mathrm{MC}$ 
of $m_x$ for a system with $L=6$ at a field of $h=0.3J$ and for two different temperatures $T=0.1J$ and $T=0.2J$ in Fig.~\ref{fig:comp_auto}. 
We observe a rather similar decay of the autocorrelation functions at $T=0.1J$, while at $T=0.2J$ the grouped plaquette algorithm even exhibits a slightly faster decay of the autocorrelations. 
Overall, both algorithms exhibit a rather similar performance, and we thus employed both for our examination of the field-induced phase transition. 
Before we present the results of our simulations, we finally describe a QMC algorithm for the TFIM on the pyrochlore lattice that is  based a decomposition  of the Hamiltonian in terms of tetrahedra, which form the  natural building blocks of the pyrochlore lattice. 

\subsection{\label{sec:tetra}Tetrahedron decomposition scheme (T-SSE)}

\begin{figure}[t]
  \centering
  \includegraphics[]{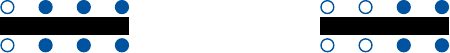}
  \caption{Different configurations of the $H_{t,2}$ operators with finite weights used in the T-SSE algorithm.
   Configurations denoted 2-2 and 3-1 contribute with weights $8J$ and $6J$, respectively.}
  \label{fig:tetrahedron_ising_operator}
\end{figure}

In this section, we describe a QMC method for the TFIM on the pyrochlore lattice, the T-SSE algorithm, which is based on a decomposition of $H$ in terms of single-site and tetrahedron terms, i.e., the basic building blocks of the pyrochlore lattice. 
A convenient partitioning of the Hamiltonian fulfills certain properties.
First, it changes the original Hamiltonian \eqref{eq:model} only up to a constant.
For the TFIM, this constant renders the weights of the SSE simulation positive in order to eliminate the QMC sign problem.
Second, it may furthermore allow one to set the weights of certain local spin configurations to zero and thus removes the need to sample these configurations.
Since we consider the antiferromagnetic TFIM model, we choose the  two fully polarized, ferromagnetic configurations on each tetrahedron to have zero weight. Based on this reasoning, we define the tetrahedron-based operator terms as follows: for a given tetrahedron $t$ from the pyrochlore lattice, 
\begin{align}
  H_{t,2}=6J-J\Big(&\sigma_{i(t)}^{z}\sigma_{j(t)}^{z}+\sigma_{i(t)}^{z}\sigma_{k(t)}^{z}+\sigma_{i(t)}^{z}\sigma_{l(t)}^{z}\nonumber\\
  +&\sigma_{j(t)}^{z}\sigma_{k(t)}^{z}+\sigma_{j(t)}^{z}\sigma_{l(t)}^{z}+\sigma_{k(t)}^{z}\sigma_{l(t)}^{z}\Big),
  \label{eq:tetrahedron_ising_operator}
\end{align}
where $i(t),j(t),k(t)$, and $l(t)$ refer to the four different sites of the tetrahedron.
The total number of tetrahedra on a finite system of linear extent $L$ with $N=4L^3$ sites is equal to $N_t=2L^3$.
The operators for the transverse-field terms in Eq.~\eqref{eq:mc_decomp_flip} and the constant operators  in Eq.~\eqref{eq:mc_decomp_transverse} are retained.
For the new operator in Eq.~\eqref{eq:tetrahedron_ising_operator}, we notice that the nonzero weights now depend on the configuration of the spins connected by the operator.
These different configurations and their respective weights are shown in Fig.~\ref{fig:tetrahedron_ising_operator}. In this vertex representation, the $H_{t,2}$ operators are represented as vertices with a total of eight legs.  
As mentioned above, the ferromagnetic states on the tetrahedron have zero weight.
All other configurations fall into two subgroups, which we denote as  2-2 configurations and 3-1 configurations.
Namely, the labels 3-1 and 2-2 refer to  the number of equally oriented spins on the tetrahedron.
The 2-2 configurations (3-1-configurations) contribute to the  weight with a factor of $8J$ ($6J$); cf. Fig.~\ref{fig:tetrahedron_ising_operator}.

As in  all previous cases, the 
 diagonal update  inserts and removes diagonal operators from the operator string with probabilities according to their weight in the partition function. Here, we explain it for the T-SSE algorithm in more detail. In particular, 
the diagonal update proceeds by traversing the operator string from propagation level $q=0$ to $q=\Lambda$. If the $q$th operator is an identity $H_{0,0}$ operator, a Monte Carlo update of the operators sequence is proposed that replaces $H_{0,0}$ by a randomly chosen diagonal Hamiltonian term. This may  either be a local operator of type $H_{i,0}$, or an Ising  operator $H_{t,2}$ on a tetrahedron of the pyrochlore lattice. 
We first decide whether the identity operator $H_{0,0}$ shall be replaced by a diagonal operator at all, with probability 
\begin{equation}
  P(H_{0,0}\rightarrow \{H_{i,1},H_{t,2}\})=\frac{\beta\left(h N + 14 J N_{{t}}\right)}{L-n}.
  \label{eq:tetrahedron_insertion_diag}
\end{equation}
Here, the factor $14J$ originates from the sum of the two weights of a tetrahedron Ising operator ($8J+6J$).
If we decide to replace $H_{0,0}$ by an operator, we next determine which operator should be inserted.
The probabilities for the two possibilities  ($H_{i,1}$ or $H_{t,2}$) are given by
\begin{align}
P(H_{i,1})=\frac{hN}{hN+14JN_{{t}}},\quad 
P(H_{t,2})=\frac{14JN_{{t}}}{hN+14JN_{{t}}}.\label{eq:tetrahedron_insertion_ising}
\end{align}
In case we decide for an on-site operator $H_{i,1}$, we randomly pick one of the $N$ sites $i$ from the pyrochlore lattice and
replace $H_{0,0}$ on propagation level $q$ by the corresponding  on-site operator $H_{i,1}$.
In case an Ising operator is chosen, we next decide whether to insert a 2-2 or a 3-1 operator.
The corresponding probabilities are fixed by the matrix elements of these operators as 
$\frac{8J}{14J}=\frac{4}{7}$ for the 2-2  and 
$\frac{6J}{14J}=\frac{3}{7}$ for the 3-1 operators.
Finally, we randomly pick one of the $N_{{t}}$  tetrahedra from the pyrochlore lattice.
If the spin arrangement of the chosen tetrahedron $t$ for the Ising operator does not match the 3-1 or the 2-2 arrangement, the insertion is rejected, while  otherwise we replace $H_{0,0}$ at level $q$ by the operator $H_{t,2}$ on the chosen tetrahedron $t$. 
If, on the other hand the operator at the propagation level $q$ is a diagonal operator (i.e., of type $H_{i,1}$ or $H_{t,2}$), we attempt to replace it by an identity operator $H_{0,0}$, according to the probability 
\begin{equation}
  P(\{H_{i,1},H_{t,2}\}\rightarrow H_{0,0})=\frac{L-n+1}{\beta\left( N h+14 J N_{{t}} \right)}.
  \label{eq:tetrahedron_update_diag_deletion}
\end{equation}
The cluster update  of the T-SSE algorithm also proceeds within the framework of the Swendsen-Wang scheme.
In order to do so, we define the cluster construction rules for the 2-2 and 3-1 operators separately.
In both cases, the number of exit legs is always even, since we treat diagonal operators.
For the same reason, we always have to exit on the entrance leg as well as the leg directly opposite to the entrance leg.
Furthermore, we cannot exit on two or six legs, since this would transform a 2-2 operator (3-1 operator) into a 3-1-operator (2-2 operator), which is forbidden when using the Swendsen-Wang scheme due to their different  weights. 

\begin{figure}[t]
  \centering
  \includegraphics[]{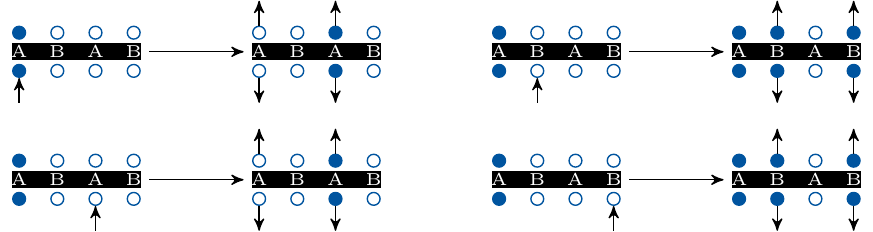}
   \caption{Illustration of the cluster building rules for the T-SSE algorithm for the 3-1 operators.
    The cluster building rules are shown by the marked entrance and exit legs of the vertices.
    The change of the local spin configuration upon an accepted cluster flip are also shown. 
  }
  \label{fig:tetrahedron_offupdate_31}
\end{figure}

We first explain the cluster construction rules for the 3-1 operators: 
for the vertex of such an operator, three spins point in one direction and the other spin points in the opposite direction.
In order to formulate a cluster construction rule, we first separate the different sites of the vertex into two groups, A and B, such that each group contains exactly two sites. This generates a total of  six  possibilities for each 3-1 operator. 
This assignment is performed randomly for all 3-1 operators that are contained within  the current operator string before starting the cluster construction.
The cluster building rule for the 3-1 operators is now rather simple: we exit on all legs that are attached to a site with the same label as the input site. 
It corresponds exactly to the rule that we already used for the G-SSE algorithm.
This rule is illustrated in Fig.~\ref{fig:tetrahedron_offupdate_31}.

\begin{figure}[t]
  \centering
  \includegraphics[]{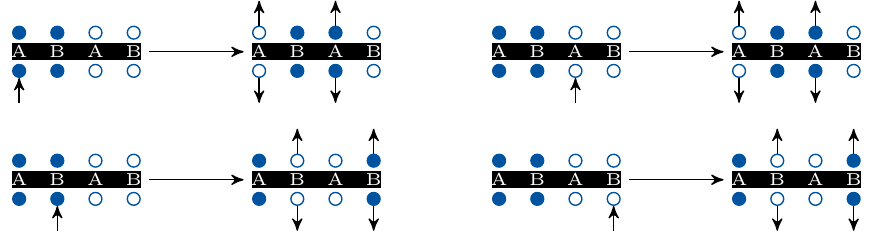}
   \caption{Illustration of the cluster building rules for the  T-SSE algorithm for the 2-2 operators.
    The cluster building rules are shown by the marked entrance and exit legs of the vertices.
    The change of the local spin configuration upon an accepted cluster flip are also shown. 
  }
  \label{fig:tetrahedron_offupdate_22}
\end{figure}

For the 2-2 operators, we can use the same overall procedure as for the 3-1 operators.
The four sites of a vertex are again assigned  to two groups, A and B. Each group contains again two sites.
But in the case of the 2-2 operators, we add an additional constraint:
each group must contain one spin pointing up and one spin pointing down.
Given this additional constraint, there are four possible assignments for the 2-2 operators.
The rule for the cluster building process is then the same as above:
we exit on all legs that are attached to a site within the same group as the site of the entrance leg.
The update rule is summarized in Fig.~\ref{fig:tetrahedron_offupdate_22}.
One Monte Carlo step consists of one diagonal and one cluster-update step.
Thus, all three algorithms (B-SSE, G-SSE, and T-SSE) can be compared directly with regards to the number of performed updates.

We compare the results obtained using the T-SSE algorithm to those from the other algorithms in Fig.~\ref{fig:comp_temp_scan_4x4x4_h_0_5}. The results obtained from the T-SSE algorithm are seen to be in accord with those using the B-SSE and the G-SSE algorithms.  In terms of  performance however, we did not observe a superior behavior of the T-SSE algorithm. An explicit comparison is shown in Fig.~\ref{fig:comp_auto}, which includes also the 
autocorrelation function from the T-SSE algorithm. For both temperatures, we observe that within the considered parameter regime near the anticipated field-induced quantum phase transition, the B-SSE and the G-SSE algorithms show better performance. From examining the typical sizes of the constructed clusters, we observe that the T-SSE algorithm tends to construct larger clusters, which  cover almost the full extent of the SSE configuration, as compared to the other algorithms. The flip of such a large cluster  does not lead to an efficient configurational update, since it  essentially corresponds to a trivial inversion of  all the spins of the SSE configuration, which does not result in  a  new configuration in terms of the physical domain structure. 
This buildup of large clusters thus degrades the performance of the T-SSE algorithm  compared to  approaches based on  smaller elementary units (bonds or triangular plaquette), for which the constructed clusters are less system-size spanning. It may be possible  to modify the T-SSE cluster-update scheme in order to overcome these difficulties, but we did not further explore this possibility. Instead, in the following section 
we present our results from applying the various algorithms to study the field-induced quantum phase transition of the TFIM on the pyrochlore lattice.

\section{\label{sec:results}  Results}

\begin{figure}[t]
  \includegraphics[]{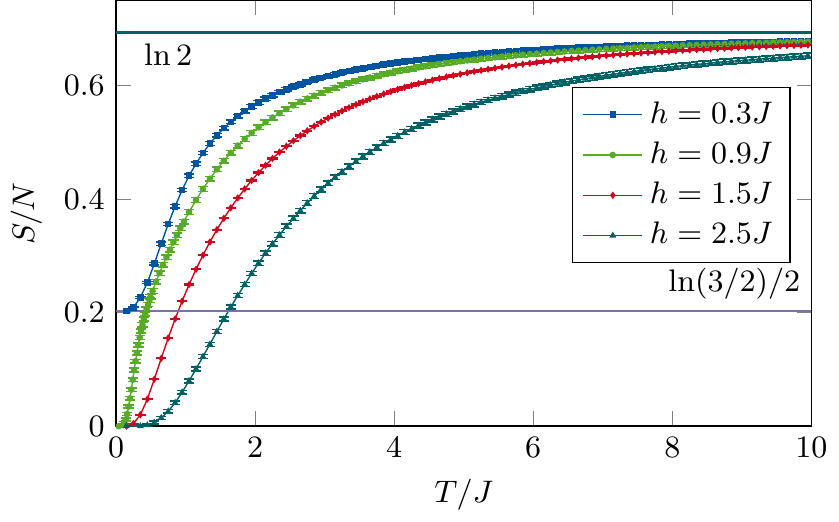}
  \caption{Temperature dependence of the thermodynamic entropy $S$ of the
TFIM on the pyrochlore lattice at $h=0.3J$ (inside the CQSL regime) and $h=0.9J$, $h=1.5J$, and $h=2.5J$ (within the HFP region)  as obtained for a $L=6$ pyrochlore lattice using the B-SSE algorithm. The upper and lower horizontal lines indicate the values of $\ln 2$ and $\ln(3/2)/2$, respectively.    }
  \label{fig:L6entropy}
\end{figure}

After having described various SSE algorithms, based on different Hamiltonian decompositions and cluster-update schemes in the previous section, we  report in this section the  results that we obtained upon applying these different algorithms to the TFIM on the pyrochlore lattice. 
Apart from those cases that are stated otherwise, all the data shown in this section are based on QMC simulations with $10^6$ thermalization steps, followed by another $10^6$ measurement steps. 

In the following, we will focus on the intermediate field region near $h_c\approx 0.6J$ of  the anticipated discontinuous quantum phase transition to examine the thermodynamic properties in this regime. To begin with however, we first demonstrate that the low-temperature thermodynamic behavior  is indeed rather different between the low- and high-field regimes: Namely, the low-field QSCSL  regime  exhibits a characteristic  finite residual entropy at low $T$, whereas the entropy vanishes for $T\rightarrow 0$ in the high-field regime. To illustrate this, we  show in Fig.~\ref{fig:L6entropy} the temperature dependence of the thermodynamic entropy $S$ of the TFIM on the pyrochlore lattice at $h=0.3J$, inside the CQSL regime,  and at $h=0.9J$, $h=1.5J$, as well as $h=2.5J$ in the HFP region, i.e. on both sides of the quantum phase transition. We obtain $S$ from a thermodynamic integration,
\begin{equation}
  S(T)= S(\infty) - \int_T^\infty\frac{C_h(T')}{T'}  dT' 
  \label{eq:entropy}
\end{equation}
of the specific heat, which we estimate from the derivative $C_h=(\partial E/\partial T)_h$ of the energy $E$ that we computed using the B-SSE scheme, and where $S(\infty)=N\ln2$ for the spin-1/2 system.  As seen in Fig.~\ref{fig:L6entropy}, the entropy in the HFP phase for $h=0.9J$, $h=1.5J$, and $h=2.5J$ vanishes for $T\rightarrow 0$, whereas, within the QCSL regime at $h=0.3J$, we observe a residual low-temperature entropy $S_0=S(T=0)$ that is in good accord with the Pauling estimate of $S_0\approx N\ln(3/2)/2$, based on the ice rule~\cite{Pauling35}. We next focus on the low-temperature behavior in the regime of the quantum phase transition 
near $h_c\approx0.6J$ for the remainder of this section. 

\begin{figure}[t]
 \centering
 \includegraphics[]{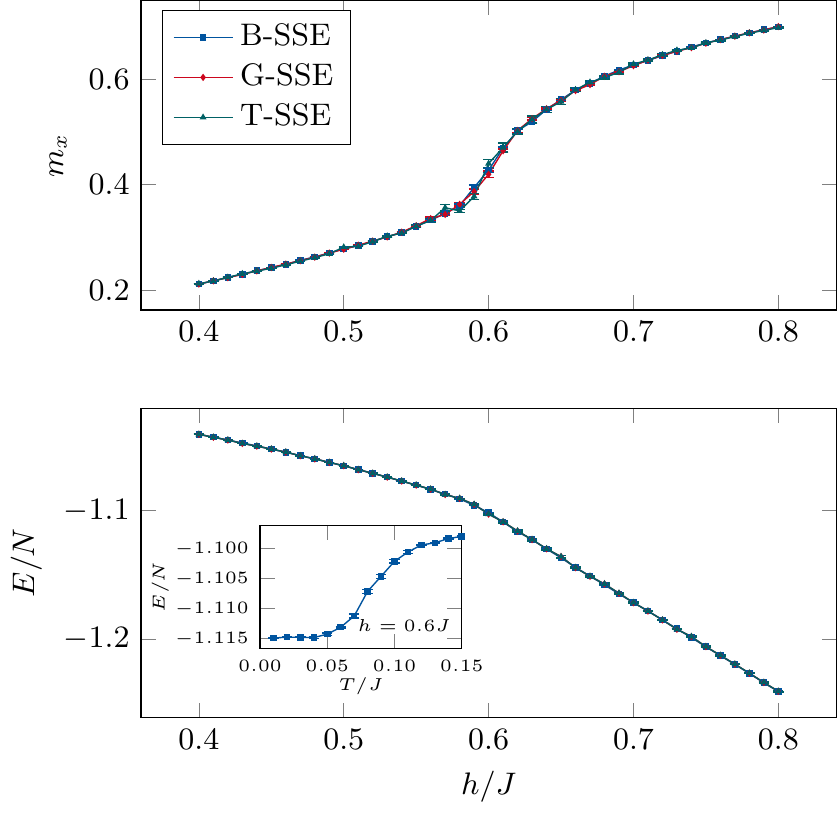}
 \caption{Transverse magnetization $m_x$ (top panel) and energy $E$ (bottom panel) of the TFIM on the pyrochlore lattice for $L=4$ as a function of the transverse-field strength $h$ at $T=0.1J$, as obtained from the different QMC algorithms.
 The inset shows the temperature dependence of the energy $E$ at a fixed transverse-field strength of $h=0.6J$ from using the B-SSE method on an $L=4$ system.}
  \label{fig:L4T0.1}
\end{figure}

Signatures of the underlying field-induced quantum phase transition of the TFIM on the pyrochlore lattice for $h\approx 0.6J$ can be observed already at a temperature of $T=0.1J$, which is considered in Fig.~\ref{fig:L4T0.1}. This figure shows  the QMC results for the energy $E$ and the transverse magnetization $m_x$ as a function of the   
transverse-field strength $h$ as obtained for a lattice with $L=4$, using the various QMC algorithms from the previous section.
Within the error bars, the results from all three algorithms provide data consistent with the T-SSE results showing larger fluctuations as compared to the other two methods, in accord with our analysis in the previous section.  
For the simulations that we performed for the remainder of this paper, we thus used the B-SSE method only. 

 At the temperature of $T=0.1J$ considered in Fig.~\ref{fig:L4T0.1}, the data for the transverse magnetization $m_x(h)$ is  consistent with a continuous $h$ dependence,
and moreover exhibits an enhanced slope near $h\approx 0.6J$, i.e., an enhanced transverse magnetic susceptibility $\chi_h= \partial m_x/\partial h$.
The transverse-field dependence of the energy $E(h$) also shows
a similarly enhanced slope at $h\approx 0.6J$ at this temperature. Correspondingly, we observe an enhanced slope in $E(T)$ in this regime, i.e., an enhanced specific heat $C_h=(\partial E/\partial T)_h$, as seen from the inset of Fig.~\ref{fig:L4T0.1}, which shows the temperature dependence of the energy $E(T)$ at a fixed transverse-field strength of $h=0.6J$.  

\begin{figure}[t]
 \centering
 \includegraphics[]{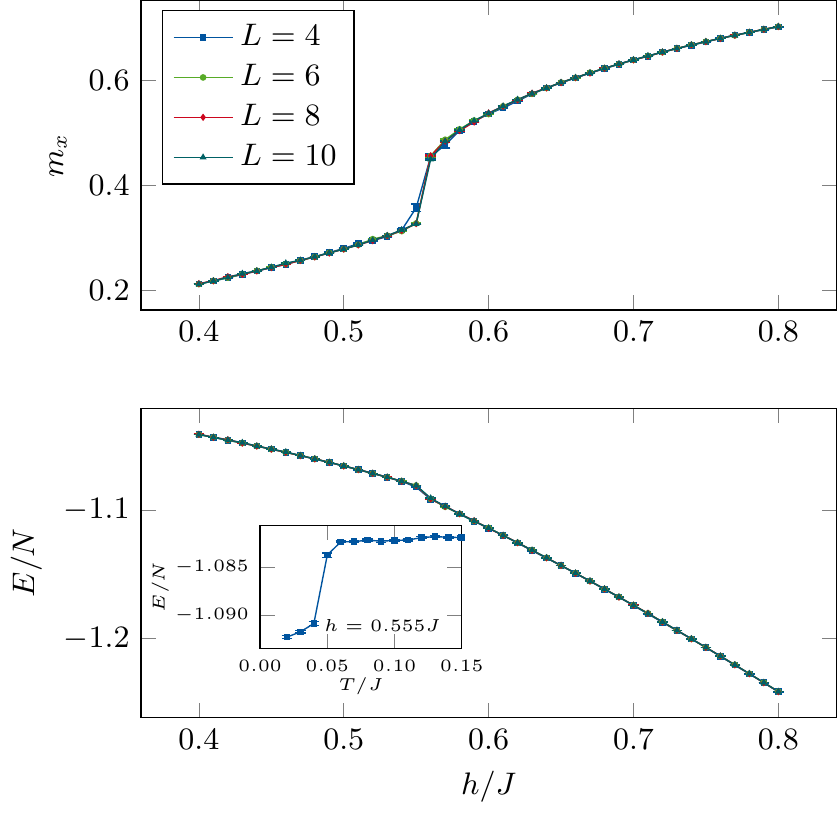}
 \caption{Transverse magnetization $m_x$ (top panel) and energy $E$ (bottom panel) of the TFIM on the pyrochlore lattice for different values of $L$ as a function of the transverse-field strength $h$ at $T=0.05J$, as obtained using the B-SSE algorithm. 
 The inset shows the temperature dependence of the energy $E$ at a fixed transverse-field strength of $h=0.555J$ on the $L=4$ system.}
  \label{fig:LvarT0.05}
\end{figure}

Upon  decreasing the temperature further down to $T=0.05J$, we observe in Fig.~\ref{fig:LvarT0.05} 
a more rapid, steplike increase in $m_x(h)$ near $h \approx 0.55J$, which is robust with respect to increasing the system size $L$. The data in Fig.~\ref{fig:LvarT0.05} thus provide indication that  discontinuous behavior emerges in the 
thermodynamic limit  at this temperature (for a finite system instead the thermodynamic quantities cannot  exhibit true discontinuities at finite $T$). 
The energy $E(h)$ similarly exhibits a kink at  $h \approx 0.55 J$ that is consistent also with a discontinuity, indicative of a finite latent heat across the $h$-driven transition line. 
This is  seen more explicitly in the jump of $E(T)$ in the inset of Fig.~\ref{fig:LvarT0.05}, which shows the temperature dependence of the energy $E(T)$ at a fixed transverse-field strength of $h=0.555J$.  

\begin{figure}[t]
 \centering
 \includegraphics[]{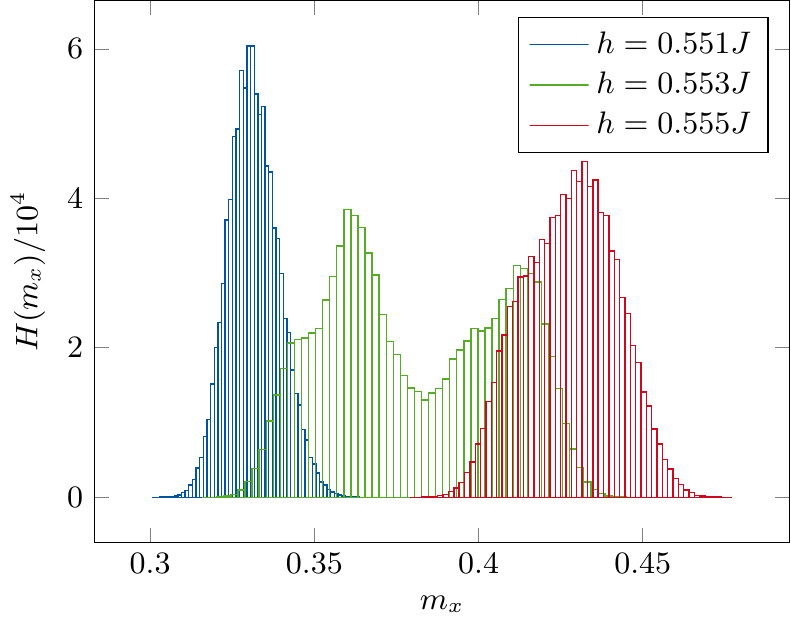}
 \caption{Histogram $H(m_x)$ of the transverse magnetization $m_x$ for the TFIM on the pyrochlore lattice for $L=8$ at $T=0.05J$ for different values of the transverse-field strength $h$ in the vicinity of the quantum phase transition, as obtained using the B-SSE algorithm.}
  \label{fig:histo}
\end{figure}

Further indications of discontinuous behavior of the transverse magnetization at  $T=0.05J$ are obtained from examining histograms $H(m_x)$ of $m_x$ for different values of the field strength. Figure~\ref{fig:histo} shows QMC histograms $H(m_x)$ 
from an $L=8$ lattice. At $h =0.551J$, the histogram $H(m_x)$ exhibits a single peak in the regime $m_x<0.38$, and for $h=0.555J$ it  shows a single peak for  $m_x>0.38$, whereas the histogram for $h=0.553J$ exhibits a two-peak structure with contributions from   configurations of both  the low- and high-$m_x$ regions. We obtained similar histograms also for the other systems sizes shown in  Fig.~\ref{fig:LvarT0.05}.   Thus, we  conclude that our QMC data at $T=0.05J$ are consistent with a low-temperature discontinuous transition in the thermodynamic limit between the low- and the high-field regime of the TFIM on the pyrochlore lattice. The critical field strength of $h_c=0.55(5)J$ that we estimate from our simulations at $T=0.05J$ falls slightly below the estimate $h_c\approx 0.6J$ that was obtained from the series expansion calculations~\cite{Roechner17}, but can be considered consistent with this value, given the uncertainties in both approaches.

\begin{figure}[t]
 \centering
 \includegraphics[]{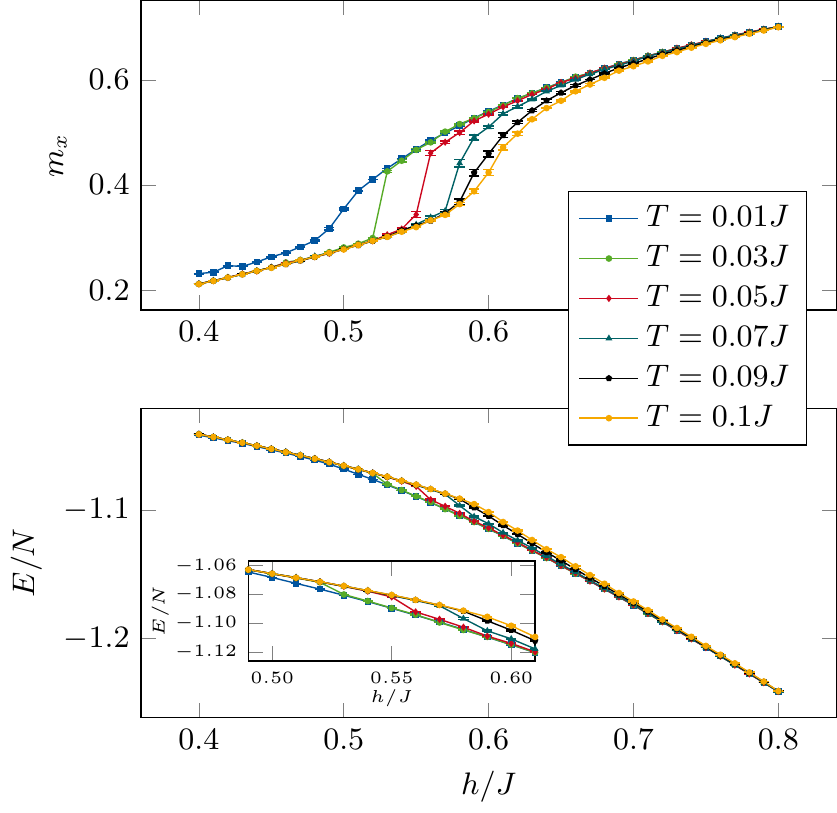}
 \caption{Transverse magnetization $m_x$ (top panel) and energy $E$ (bottom panel) of the TFIM on the pyrochlore lattice for $L=4$ as a function of the transverse-field strength $h$ at different temperatures $T$, as obtained using the B-SSE algorithm. The inset in the bottom panel focuses on the transition region around $h=0.55J$.}
  \label{fig:L4Tvar}
\end{figure}

We finally examine the $h$ dependence of $E$ and $m_x$ upon further varying the temperature $T$. For this purpose, we performed QMC simulations for finite-size systems with $L=4$ 
down to the lowest accessible temperatures of $T=0.01J$. In order to ensure a better sampling 
for $h\approx 0.5J$, we performed the statistical averaging based on a set of ten independently thermalized simulations at $T=0.01J$ for $h$ between $0.45J$ and $0.55J$, thereby increasing  the total number of Monte Carlo measurement steps by a factor of 10 for this parameter set. 
The results from the QMC simulations are summarized in Fig.~\ref{fig:L4Tvar}.
In accord with the data shown in Fig.~\ref{fig:L4T0.1}, we observe an essentially continuous $h$ dependence  of $E$ and $m_x$  
for temperatures $T\gtrsim0.07J$. 
At lower temperatures, $0.01J \lesssim T\lesssim0.07J$, the 
data for $m_x$ in  
Fig.~\ref{fig:L4Tvar} 
instead show a  steplike increase in the $h$ dependence. The value of the transverse field at the position of the step in $m_x$ furthermore decreases upon lowering the temperature $T$. 

However, as we already mentioned in Sec.~\ref{introduction}, the ground state of the TFIM on a finite pyrochlore lattice does not exhibit a level crossing transition, so that  for a finite system the transverse magnetization $m_x$ (in addition to $E$) evolves continuously upon increasing $h$ at $T=0$.  A discontinuous quantum phase transition may occur in the thermodynamic limit from the closing of the excitation gap $\Delta$ at $h_c$, while the gap $\Delta$ remains nonzero on a finite system for any value of $h$.  
For temperatures $T$ above $\Delta$, one may  observe behavior   similar to   the  thermodynamic limit, e.g., a developing discontinuity, also on large, finite lattices. However, at temperatures well below $\Delta$, we expect the $h$ dependence of, e.g., $m_x(h)$, to be continuous on a finite system. This continuous behavior is of course more readily resolved on small lattices with  a larger minimum gap. 
In accord with this expectation, we observe in Fig.~\ref{fig:L4Tvar} that  on the $L=4$ system 
the transverse magnetization $m_x$  shows a smoother  $h$ dependence at $T=0.01J$ as compared to the more steplike  increase of the data for, e.g.,  $T=0.05J$. 
While we cannot exclude based on our data that the observed behavior of $m_x$ results due to other reasons, such as, e.g., a further low-temperature phase between the CQSL and the HFP regime, 
 we also did not obtain any clear indication in favor of such alternative scenarios (such an intermediate phase between the low- and high-field regimes has, however, been reported recently for the TFIM on the checkerboard lattice -- a two-dimensional system of coupled tetrahedra~\cite{Henry14}). Our QMC data are thus  in good qualitative and also quantitative agreement with the conclusions from Refs.~\onlinecite{Savary17,Roechner17}, suggesting a discontinuous quantum phase transition of the TFIM on the pyrochlore lattice.

\section{\label{sec:conclusions} Conclusions}

We examined the low-temperature properties of the TFIM on the pyrochlore lattice using various QMC algorithms  formulated within the SSE  approach. For this purpose, we considered several previously proposed cluster-update schemes, based on either bond or triangular plaquette decompositions of the model Hamiltonian. We furthermore presented an adapted version of the plaquette-based cluster-update scheme (the grouped plaquette method), which avoids the convergence problem that we identified for  the plaquette algorithm in the original formulation. We also proposed a tetrahedron-based  algorithm for the pyrochlore lattice. Within the considered parameter regime of the TFIM, i.e., near the field-induced quantum phase transition, we found the bond and the grouped plaquette algorithms to perform more efficiently in order to extract the thermodynamic properties on systems with several hundred spins and  down into the low-temperature regime.  

Our QMC results for the low-temperature behavior 
support  the  scenario in Refs.~\onlinecite{Savary17,Roechner17} in terms of a discontinuous quantum phase transition of the TFIM on the pyrochlore lattice that separates the low-field CQSL from  the HFP regime. Within this scenario, our data furthermore indicate that the discontinuous transition between the low- and high-field regimes persist also beyond the zero-temperature limit up to at least $T=0.05J$, whereas  for even larger temperatures, $T\gtrsim 0.1J$, the thermodynamic quantities evolve smoothly with the transverse-field strength. 
A natural interpretation of these findings is as follows: the discontinuity of the  quantum phase transition extends as a finite-temperature first-order phase transition line up to temperatures of $T\approx 0.1 J$, and eventually terminates in a critical point, i.e., similar to the well-known liquid-gas transition line~\cite{rcdlt}. In fact, 
within both the low- and high-field regimes, the TFIM on the pyrochlore lattice does not feature any long-range order from spontaneous symmetry breaking, since the strong geometric frustration of the pyrochlore lattice hinders the formation of a long-ranged ordered state. This absence of symmetry breaking on both sides of the phase transition line resembles the liquid-gas system, and, based on this analogy,  we expect the critical point that terminates the first-order coexistence line to belong to the three-dimensional Ising universality class. By the same analogy,  the transverse magnetization corresponds to the density in the liquid-gas system. We mention here that a scenario similar to the one outlined above has  been identified recently in a strongly frustrated two-dimensional Heisenberg quantum spin model~\cite{Stapmanns18}, for which the absence of finite-temperature symmetry breaking is enforced already by the Mermin-Wagner theorem~\cite{Mermin66}. 

It would be interesting to further assess the above scenario of an extended first-order transition line with a critical point for the TFIM on the pyrochlore lattice. However, based on our analysis we conclude that for this purpose further improvements of the QMC methods are imperative in order to access the range of system sizes that are required for a reliable finite-size scaling analysis. This is needed for  an accurate tracing of the first-order transition line and to locate and eventually identify its termination point. 

\section*{Acknowledgments}
We thank Sounak Biswas and Kedar Damle for sharing their insights on quantum Monte Carlo algorithms for the transverse-field Ising model on frustrated lattices and Tommaso Roscilde for valuable  discussions on the membrane algorithm of Ref.~\onlinecite{Henry14} and the phase diagram of the transverse-field Ising model on the checkerboard lattice.   We furthermore  acknowledge support by the Deutsche Forschungsgemeinschaft (DFG) under Grants No. FOR 1807 and No. RTG 1995. Furthermore, we thank the IT Center at RWTH Aachen University and the J\"ulich Supercomputing Centre for access to computing time through JARA-HPC.

\end{document}